\documentclass[journal]{IEEEtran}
\usepackage{graphicx}
\newcommand{\ie}{{\em i.e.,}\ }
\newcommand{\eg}{{\em e.g.,}\ }
\newcommand{\et}{{\em et al.}\ }
\newcommand{\NP}{{\em NP}}

\newcommand{\x}{{\bf x}}
\newcommand{\p}{{\bf p}}

\def\cornerbox#1#2#3{\setbox1=\hbox{#1}
\dp1=0pt\par\hangindent\wd1\hangafter-#2\noindent
\hskip-\wd1\raise#3\box1\ignorespaces}

\begin{document}

\title{Foundations of Swarm Intelligence:\\ \LARGE From Principles to
Practice}

\author{Mark Fleischer\\
Institute for Systems Research\\ University of Maryland \\ College
Park, Maryland 20742}

\maketitle

\markboth{Swarming: Network Enabled C4ISR\hfill\copyright\ 2003 by
Mark Fleischer. All rights reserved.\hspace{15pt}}{}

\begin{abstract}

Swarm Intelligence (SI) is a relatively new paradigm being applied
in a host of research settings to improve the management and
control of large numbers of interacting entities such as
communication, computer and sensor networks, satellite
constellations and more. Attempts to take advantage of this
paradigm and mimic the behavior of insect swarms however often
lead to many different implementations of SI. The rather vague
notions of what constitutes self-organized behavior lead to rather
{\em ad hoc} approaches that make it difficult to ascertain just
what SI is, assess its true potential and more fully take
advantage of it.
This article provides a set of general principles for SI research
and development. A precise definition of {\em self-organized
behavior} is described and provides the basis for a more axiomatic
and logical approach to research and development as opposed to the
more prevalent {\em ad hoc} approach in using SI concepts.

The concept of {\em Pareto optimality} is utilized to capture the
notions of efficiency and adaptability. A new concept, {\em Scale
Invariant Pareto Optimality} is described and entails symmetry
relationships and scale invariance where Pareto optimality is
preserved under changes in system states. This provides a
mathematical way to describe {\em efficient tradeoffs of
efficiency} between different scales and further, mathematically
captures the notion of the {\em graceful degradation of
performance} so often sought in complex systems.



\end{abstract}


\section{Introduction}
\label{sec:introduction}
\PARstart{T}{oday's} communications networks have become
enormously complex systems.  New technologies from sensor
networks, web-enabled PDAs, remote surgery systems to
constellations of orbiting satellites all require enormous numbers
of communicating and interacting entities.  These entities must
work together harmoniously to be effective. As the numbers of
these interacting entities increases, ensuring their efficient
operation becomes increasingly difficult. Indeed, for the past
three decades this growth in general has approximately doubled
every 18 months \cite[p.32]{Leon-Garcia00}. New paradigms of
modern warfare also indicate an accelerated growth in the numbers
of interacting systems.  Amidst this growth, there is a growing
consensus among experts that current network management approaches
will be insufficient to handle the level of complexity that is
envisioned\footnote{As stated by the National Research Council in
a report on countering terrorism: ``Research is also needed for
self-adaptive networks that can reconfigure themselves in response
to damage and changes in demand, and that can degrade
gracefully.''\cite{NRC02}} \cite{NRC02}. Consequently, new
approaches for network management and control in complex systems
are needed.

One promising approach is based on what is often referred to as
{\em Swarm Intelligence} (SI).  The term SI has come to represent
the idea that it is possible to control and manage complex systems
of interacting entities even though the interactions between and
among the entities being controlled is, in some sense, minimal.
This notion therefore lends itself to forms of distributed control
that may be much more efficient, scalable and effective for large,
complex systems.

The underlying features of SI are based on observations of social
insects.  Ant colonies and beehives, for example, have the
interesting property that large numbers of them seem to conduct
their affairs in a very organized way with seemingly purposeful
behavior that enhances their collective survival. Surprisingly and
paradoxically, these insects seem to utilize very simple rules of
interaction. This phenomenon is very similar to those addressed in
other domains of inquiry involving {\em complexity} such as
cellular automata and the study of chaos
\cite{Chaudhuri97,Pound85,Prigogine84}. These areas along with SI
have perplexed a large number of scientists for many years
\cite{SI-Bonabeau99}. How is it that ``swarms'' of creatures with
relatively low brain power and communications capabilities can
engage is what is often termed ``emergent behavior'' reflective of
some ``collective intelligence'' \cite[p.6]{SI-Bonabeau99}
---behavior that seems to exhibit a more global purpose?

Unfortunately, there is no widely agreed upon definition of what
SI is or how it should or could be mathematically defined or
characterized.  Many terms have been associated with SI such as
{\em emergent behavior}, {\em self-organized behavior}, {\em
collective intelligence}, and the like and have been used in a
variety of contexts and associated with a host of applications
\cite{Bull1997,Mani1994,Scho1996}, but these terms also suffer
from vague definitions or descriptions. There is no general,
mathematically oriented description that ties all of these
concepts together.

The lack of precise definitions and, hence, theoretical
foundations, poses a number of significant problems and even
causes confusion. The lack of precise definitions is the least of
the problems---this confusion also entails missed opportunities as
well.  All these different descriptions and implementations muddy
the waters of how to productively utilize SI concepts. Without a
clear understanding of what SI is and how and why it arises, it is
very difficult to envision how to take advantage of its true
potential.

The remedy for this apparent confusion comes from new perspectives
that illuminates the {\em fundamental\/} properties of SI. This
article seeks to do just that by articulating some useful ideas
based on perspectives from evolution, notions of efficiency, and
adaptability coupled with a more formal definition of
self-organized behavior.

This article formulates three strategic or foundational frameworks
from which to build a successful theory of SI and which provide
guidelines for how to implement SI concepts. These frameworks are
based on the development of the successful theories pertaining to
the {\em simulated annealing} (SA) algorithm. The application of
these frameworks to SI may provide the necessary yet still missing
ingredients for developing a better understanding of SI. This may
lead to entirely new ways of viewing and understanding this
paradigm and ultimately allow for more practical implementation
schemes.

The most important part of this foundational triad, and a
necessary ingredient for developing a solid theoretical foundation
for SI, is the articulation of some {\em first principles} based
on the relevant laws of nature and their implications. This serves
as a guide that governs and constrains the articulation of the
other important components in developing a useful theory. For SI,
these first principles are based on the laws of evolution and
natural selection.  As the reader will discover, its reasonable
implications suggest a more precise and mathematically useful
definition of self-organized behavior---{\em the evolution of
system states along a Pareto optimal frontier}.

The second component of this triad is the articulation of an
appropriate {\em dynamical framework}, a way of characterizing
system dynamics that is a consequence of or constrained by the
first principles. The dynamical framework suggested here is based
on a new concept in the context of SI---{\em Scale Invariant
Pareto Optimality} (SIPO). SIPO is a powerful concept that
captures notions of symmetry and scale invariance and can address
the issues of how swarms of entities communicate, modify their
behavior, and {\em adapt\/} to changing environmental
conditions---one of the hallmarks of SI. This dynamical framework
also provides a set of rules that, in effect, imposes constraints
on a system's dynamics so as to maintain consistency with the
implications of the relevant laws of nature as articulated in the
first principles.

Finally, the third component of this triad is the articulation of
an appropriate {\em problem framework}.  This provides useful ways
to abstract these ideas and allows them to be implemented. It
provides a concrete way of defining problems that help to further
narrow the issues and focus research and development efforts. This
article describes a general test-bed approach using the concept of
{\em swarming finite-state machine} (SFSM) models.

Together, this triad of frameworks, or {\em meta-formalism},
present a unified scheme for approaching the research problems and
investigating ways to implement SI concepts. It addresses how
swarms of entities must communicate and modify their behavior in
response to information from other entities and their environment
for there to exist the emergent, self-organized behavior known as
``swarm intelligence''. This set of perspectives leads to a more
precise mathematical definition of SI, describes ways to more
fully take advantage of this paradigm and, constructively address
any of its inherent limitations.

The rest of this article is organized as follows:  Section
\ref{sec:background} provides more detail on the current research
environment regarding SI and its myriad of applications.  Section
\ref{sec:researchstrategy} describes why the three formalisms
described above provide a sound basis for developing the
theoretical foundations of SI by describing similarities to the
{\em de facto} frameworks associated with simulated annealing.
Sections \ref{sec:evolution} through \ref{sec:problemframework}
describe the three frameworks on a conceptual level. Finally,
Section \ref{sec:conclusion} provides concluding remarks.

\section{Background: The Swarm Intelligence Paradigm}\label{sec:background}
\subsection{Observations of Social Insects}
Observations of social insects such as ants and ant colonies
provide a great deal of insight into their behavior and SI in
general. Ants and ant colonies have several ways of solving
different but related problems.  The main mechanism for solving
them is through the use of chemical substances known as {\em
pheromones} which have a scent that decays over time through the
process of evaporation \cite[p. 26]{SI-Bonabeau99}. These
pheromones form the basis of what amounts to a clever, and
apparently simple, communications and information storage and
retrieval system. Since pheromone strength or intensity decays
over time, it also provides a very simple information processing
mechanism that can implement forms of positive and negative
feedback \cite[pp. 9-10, 41]{SI-Bonabeau99} and {\em reinforcement
learning} mechanisms \cite[p.96]{SI-Bonabeau99}. This
``processing'' capability is illustrated in the simplicity of how
ants utilize and respond to pheromones.

As an example, consider how ants actually solve shortest path
problems.  Their motivation for solving these problems stems from
their need to find sources of food.  Efficiency dictates that they
find sources closest to their colonies.  Ants (many ants) first
set out in search of a food source by randomly choosing
(apparently randomly) several different paths. Along the way they
leave traces of pheromone \cite[p. 42]{SI-Bonabeau99}. Once ants
find a food source, they retrace their path back to their colony
(and in so doing inform other ants in the colony) by following
their scent back to their point of origin. Since many ants go out
from their colony in search of food, the ants that return first
are presumably those that have found the food source closest to
the colony or at least have found a source that is in some sense
more accessible. In this way, an ant colony can identify the
shortest or ``best'' path to the food source \cite{SI-Bonabeau99}.

The cleverness and simplicity of this scheme is highlighted when
this process is examined from what one could conceive of as the
ants' perspective---they simply follow the path with the strongest
scent (or so it seems).  The shortest path will have the strongest
scent because less time has elapsed between when the ants set out
in search of food and when they arrive back at the colony, hence
there is less time for the pheromone to evaporate. This leads more
ants to go along this path further strengthening the pheromone
trail and thereby reinforcing the shortest path to the food source
and so exhibits a form of reinforcement learning
\cite{SI-Bonabeau99,Kaelbling96,Littman93}.

But this simple method of reinforcement or positive feedback also
exhibits important characteristics of efficient group behavior.
If, for instance, the shortest path is somehow obstructed, then
the second best shortest path will, at some later point in time,
have the strongest pheromone, hence will induce ants to traverse
it thereby strengthening this alternate path. Thus, the decay in
the pheromone level leads to {\em redundancy}, {\em robustness}
and {\em adaptivity}, \ie what some describe as {\em emergent}
behavior \cite{SI-Bonabeau99}.

Many optimization algorithms attempt to imaginatively capture some
notion of SI. Indeed, many difficult optimization problems have
been solved by so-called {\em ant algorithms} such as the
Traveling Salesman Problem, the Quadratic Assignment Problem and
other \NP-hard optimization problems (see \cite{SI-Bonabeau99} for
a large number of examples and citations).  These algorithms
generally utilize some analogue of pheromone or some simple
stigmergic signalling mechanism.  AntNet \cite{DiCa1997_2} for
example, uses reinforcement learning to increase the probabilities
of using certain routes in a routing algorithm.  The probability
value is used as an analogue to pheromone.  Another example is in
\cite{Parunak02} which uses a similar update mechanism to control
unmanned aerial vehicles. These different approaches all try to
take advantage of how social insects seem to function. These
attempts to implement some SI characteristic however often are
forced to creatively sidestep the concept of self-organization and
its implications.

\subsection{The Mystery of
Self-Organization}\label{sec:self-organization} Although much has
been learned from observations of these social insects, there is
no widely agreed upon definition of what constitutes SI. Indeed,
the term {\em SI} is bandied about so often and in such a wide
variety of contexts that it causes confusion leading to many
different interpretations and implementations for various
problems. Although many of these implementations reflect some
notion of SI, they often entail other paradigms as indicated above
such as {\em reinforcement learning}
\cite{Littman93,Subramanian00} and  {\em stigmergy} which refers
to complex {\em indirect} interactions based on simple signalling
systems \cite[p.14]{SI-Bonabeau99}. Where does one paradigm end
and the other begin? Still, other research seems to freely use the
term SI when there simply are large numbers of interacting
entities (see \cite{Langton94} for a complete discussion on
competing paradigms). Does that alone suffice to describe or
define SI? Is it merely a way of somehow taking advantage of
parallel computing methods?

Notwithstanding the many different descriptions offered by many
researchers, the main features of SI seem to involve forms of
limited or minimal communications and/or interactions, large
numbers of interacting entities with limited reach, and some
globally efficient, emergent or {\em self-organized} behavior
\cite[p.9]{SI-Bonabeau99}. Bonabeau \et
\cite[p.9-11]{SI-Bonabeau99} suggests that the central features of
SI are based on the manifestation of {\em self-organization} that
arise from the interplay of four basic ingredients: 1) forms of
positive feedback, 2) forms of negative feedback, 3) the {\em
amplification of fluctuations} that give rise to structures, and
finally 4) multiple interactions of multiple entities.

But even this characterization provides very little insight into
what SI is except in very descriptive terms.  For example, it does
not fully explain in a unifying way why or how pheromones evolved
in the way they did, or why they should have the evaporative
properties they have, or why they have their chemical makeup, or
how ants can somehow distinguish among different types of
pheromone (presumably in order to identify ants from other
colonies) \cite[p.156]{Wiener61}. Pheromones are complex chemical
signalling systems, yet most of the research that deals with them
or models their effects use the concept in very limited ways, \eg
as a scalar in ant algorithms \cite{Subramanian00,AntNet} as
opposed to a more complex scheme represented by vectors. Although,
as we shall see, even simple scalars can possess enough
information related to a measure of efficiency, pheromones are
likely to have more complicated properties than their mere
intensity \cite{SI-Bonabeau99}. Indeed, Wiener \cite{Wiener61} in
his ground-breaking book {\em Cybernetics} emphasizes the
importance of intercommunication among the entities in question:
\begin{quote}
How then does the beehive act in unison, and at that in a very
variable, adapted, organized unison?  Obviously, the secret is in
the intercommunication of its members\ldots This
intercommunication can vary greatly in complexity and content
\ldots the value of a simple stimulus, such as an odor, for
conveying information depends not only on the information conveyed
by the stimulus itself but the whole nervous constitution of the
sender and the receiver of stimulus as well
\cite[p.156-7]{Wiener61}.
\end{quote}
However SI is described, one of its central characterizations is
that of {\em self-organization}.  But this also begs the question
of what constitutes SI because there is no clear understanding of
what self-organization or emergent behavior is! These terms have
been around for some time and their definitions have been and
probably will continue to be debated for some time.

Serra \et describes the concept of self-organization generally as
``highly organized behaviour even in the absence of a pre-ordained
design.'' \cite[p.1]{Serra90} (but what is `organized'?).  They go
on to further describe examples such as the resonance phenomenon
in lasers, and in cellular automata where ``unexpected and complex
behaviours [] can be considered as self-organized.''
\cite[p.2]{Serra90}.

Earlier, Prigogine\footnote{Prigogine won the Nobel Prize in 1977
for his work on the thermodynamics of non-equilibrium system.} \et
\cite{Prigogine84} described self-organization in chemical
reactions and thermodynamic contexts. His notion is quite
enlightening and emphasizes the element of {\em fluctuations} in
far-from-equilibrium system states. Non-linear system changes,
fluctuations as he refers to them, due to either random events or
chaotic dynamics, are then amplified by positive feedback
mechanisms.  This results in structures that emerge spontaneously
which often are presented as examples of self-organization. It is
interesting that Prigogine provides an example of SI based on the
clustering behavior of termites in constructing termite nests
\cite[p.181-186]{Prigogine84}.  See also
\cite[p.207-234]{SI-Bonabeau99}.  Even earlier, Wiener
\cite{Wiener61} used the term in describing the brain waves of
humans.  See \cite[Chap. 10: Brain Waves and Self-Organizing
Systems]{Wiener61}.  Again, the notion of what self-organization
is remains unclear.  Is it merely some structure or pattern?  And
if so, how is one to distinguish it from merely random effects?

What is needed to truly take advantage of SI is more than the mere
descriptions of the attributes of SI and self-organization.
Something amenable to formal mathematical definition would be
quite valuable.  This article offers a new and potentially useful
working definition of emergent or self-organized behavior for SI
based on the imperatives of evolution and natural selection (as
opposed to the imperatives of thermodynamics and irreversibility
that elicited Prigogine's definition \cite{Prigogine84}).  The
concept of self-organization therefore plays a central role in the
development of the foundational principles of SI.

These principles are founded on the notion that self-organized
behavior of the type observed in entities subject to the laws of
evolution involves forms of efficiency in resource allocation.  It
emphasizes the importance of {\em signals} as mechanisms of change
in system states subject to the imperatives of the evolutionary
pressures of natural selection.  It is important to recognize that
the response of insect swarms to external stimuli is governed by
processing systems that have been heavily influenced and affected
by the forces of evolution. This allows consideration of a much
richer spectrum of behaviors, both simple and complex, than those
implied by the mere application of positive and negative feedback
mechanisms to simple signals. In fact, the type of
self-organization described here can be framed as {\em a form of
symmetry} in that changes in the system's behavior and operating
points nonetheless leave unchanged certain attributes associated
with these operating points, namely, their Pareto optimality
\cite{Rosen95}. In short, a novel definition of self-organization
presented here is {\em system behavior that maintains its
operating points on or near a Pareto optimal frontier.} This
notion of efficiency constitutes a central feature of the
foundational principles for SI and is described in greater detail
in the next section.

\section{The Importance of Basic Principles}\label{sec:researchstrategy}

Developing a viable set of basic principles underpinning SI
concepts is essential for successful implementations. The
principles described here have a certain credibility and hence
viability because they are based on how other successful theories
appear to have been developed. This requires some discernment and
creative articulation of the main features of how these other
theories were, in fact, developed and how these features could be
applied in the case of SI. The simulated annealing (SA) algorithm
is a good example of this process and provides clues on how to
articulate the corresponding components of the central principles
underlying SI.

A great deal of theoretical results on SA have been published in
many papers and books (see \eg
\cite{Aarts,Bohachevsky,Mitra,Fleischer_Scale}). But this
algorithm was not developed or analyzed in a vacuum. Its
development was, in fact, based on the confluence of certain
factors. Three main factors or components of the research
evolution are discernable. The first component consists of a set
of {\em first principles} based on the {\em relevant laws of
nature and their implications}. The second component is an
appropriate {\em dynamical framework}. Finally, the third
component is an appropriate {\em problem framework}. These three
components form the core of what can be referred to as a set of
foundational frameworks. To see how these components provide
guidance in developing a viable research strategy for SI, an
understanding of how they worked together in the case of SA is
helpful.

In SA, the first principles were based on the laws of
thermodynamics.  The most salient implication of this law was that
the entropy of a thermodynamic system, the randomness of its
states, tends to its maximum value at any given temperature.
Metropolis \cite{Metropolis} utilized this idea in developing a
method for simulating the evolution of a complex thermodynamic
system. The underlying law of thermodynamics was manifest in the
objective function of a non-linear program (NLP). The objective of
this NLP was to maximize the entropy of a stationary Markov chain
subject to certain constraints (such as the the constraint that
expected energy level is proportional to temperature)
\cite{Bohachevsky}.

The solution of this NLP lead to an expression for the stationary
probability distribution (the Boltzmann Distribution) of each
system state \cite{Bohachevsky}.  Thus, the implication of the
first principles was the expression of the Boltzmann Distribution.
What remained was to somehow develop an appropriate mechanism that
described how this thermodynamic system changes from one state to
another while at the same time remaining consistent with the
maximization of entropy, \ie the Boltzmann Distribution, as
dictated by the fundamental laws of thermodynamics.

Thus, in addition to this fundamental law, a {\em dynamical
framework} was necessary to address the problem of how to model a
system that undergoes some type of change. The well-known
mathematical framework of Markov chains and its global and
detailed balance equations provided those necessary elements that
mathematically describe the stochastic nature of the system as it
changes from one state to another while still obeying the
fundamental laws of thermodynamics \cite{Aarts,Mitra,Metropolis}.
It was this utilization of the detailed and global balance
equations that ultimately lead Metropolis \et to his now famous
{\em Metropolis Acceptance Criterion}, the main result of
Metropolis' contribution. In effect, Metropolis `engineered' a
transition probability to fit within the framework of global and
detailed balance equations given the stationary probability, \ie
the Boltzmann Distribution.

Once these transition probabilities were defined, the mathematical
description of how a large, thermodynamic system evolves from one
state to another in a computer simulation was complete
\cite{Metropolis}.  Once this method of simulating a thermodynamic
system was in place (in 1953), the idea of reducing temperature
slowly to ``anneal'', conceived by Kirkpatrick
\cite{Kirkpatrick83} some 30 years later, was the final element
that lead to the SA algorithm.  This algorithm has since been
fully mathematically described and analyzed by a host of
researchers\footnote{Although the algorithm seems to have been
fully analyzed, this proposer believes there are still a number of
fundamental theoretical elements that await discovery.  See \eg
Fleischer \cite{Fleischer_Scale} for a description of scale
invariant properties of the SA algorithm. These properties
illustrate some potentially important symmetries in SA and may
play a role in research involving a group theoretic approach for
modelling SI.} (see \cite{Aarts,Mitra,Metropolis,Kirkpatrick83})
using various {\em combinatorial optimization problems} (COPs) as
the {\em problem framework}---a general way of characterizing a
whole host of discrete optimization problems. Thus, the dynamical
and problem frameworks provided the mathematical constraints and
research focus in a way that was consistent with the laws of
thermodynamics.

To summarize how these three components of the foundational
principles, the meta-formalism, describes the evolution of SA
theory, the first principles were based on thermodynamics manifest
as the maximization of the entropy in an NLP that lead to the
definition of the Boltzmann Distribution.  The dynamical framework
was based on the global and detailed balance equations of Markov
chains which lead to the Metropolis Acceptance Criterion and the
transition probabilities in SA. Finally, the problem framework was
based on COPs which permitted the application of the ideas onto
test bed problems for experimentation and analysis. A successful
and useful set of formalisms can therefore be modelled on this
developmental process\footnote{We believe that other theories such
as those associated with genetic algorithms (GAs) offer a similar
developmental model.} hence ought to involve first principles
based on fundamental laws, an appropriate dynamical framework, and
an appropriate problem framework.  Each of these three components
can be recast to within the context of SI. This is described in
more detail in the next three sections.

\subsection{First Principles: The Laws of
Evolution}\label{sec:evolution} Any successful theory requires
some reasonable and widely accepted assumptions on which the
theory is based. Because the SI paradigm is based on observations
of social insects, \ie living creatures, the first principles
ought to based on, involve, or at least be consistent with, the
laws of evolution and its implications. Basically, this theory
states that species evolve over generations constantly improving
their fitness to survive through the process of natural selection
and genetic mutation. Indeed, this theory forms the basis of GAs
and its associated theoretical results \cite{Reeves93,Holland75}.
But this theoretical statement by itself seems insufficient to
establish foundational principles for SI. Some {\em reasonable
implications} of this theory are needed.

One reasonable and important implication of the theory of
evolution as it relates to swarms of insects is that it selects
out those systems that exhibit {\em efficient behavior on many
scales}. This notion of efficiency is not only reasonable, but is
also based on observations of social insects and how they {\em
adapt\/} to changing environmental conditions (see
\cite{Holland75}). Indeed, it seems basic that an efficient
allocation and use of resources provides a distinct survival value
to any species.  Thus, a reasonable implication of evolution
regarding the behavior of social insects is that it lead to
behaviors that are, in some measure, {\em efficient\/}. These
notions of efficiency and adaptability have an important
mathematical formalism based on the concept of {\em Pareto
optimality}.

\subsubsection{Efficiency Via Pareto Optimality}\label{sec:Pareto}
Optimization problems are ubiquitous in the real world and social
insects must also deal with a variety of them if they are to
survive. Certainly, the efficient allocation of resources present
problems where some goal or objective must be maintained or
achieved.  Such goals or objectives are often mathematically
modelled using {\em objective functions}, functions of decision
variables or parameters that produce a scalar value that must be
either minimized or maximized. The challenge presented in these
often difficult problems is to find the values of those parameters
that either minimize or maximize, \ie optimize, the objective
function value subject to some constraints on the decision
variables.

Most complex systems, however, have {\em several\/} objectives
that must be considered when assessing overall system performance.
Operations of these systems sometimes leads to conflicting
objectives where one objective must be {\em efficiently} ``traded
off'' for another. In these {\em multi-objective optimization
problems} (MOPs) system efficiency in a mathematical sense is
often based on the definition of {\em Pareto optimality}—--a
well-established way of characterizing a {\em set\/} of optimal
solutions when several objective functions are involved (see \eg
\cite{Fleischer_EMO} and the citations therein).

In this perspective, each operating point or vector of decision
variables (operational parameters) produces several objective
function values corresponding to a single point in objective
function space (this implies a vector of objective function
values). A {\em Pareto optimum} corresponds to a point in
objective function space with the property that when it is
compared to {\em any} other feasible point in objective function
space, at least one objective function value (vector component) is
superior to the corresponding objective function value (vector
component) of this other point. Pareto optima therefore constitute
a special subset of points in objective function space that lie
along what is referred to as the {\em Pareto optimal
frontier}—--the set of points that together dominate (are superior
to) all other points in objective function space
\cite{Fleischer_MOPO}.  See also \cite{Coello01} and
\cite{Menczer00} for descriptions of multi-objective GAs (their
intriguing titles notwithstanding do not encompass the ideas put
forth here on SI).

\begin{figure}[htb]
\vspace{-.2in}
\begin{center}
\includegraphics[width=2in]{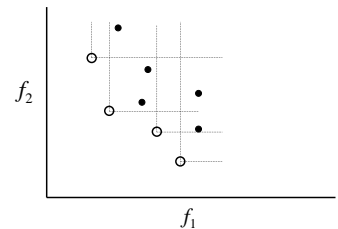}
\caption{\small The Pareto Optimal Frontier}
\label{fig:paretofrontier}
\end{center}
\end{figure}
An important characteristic of Pareto optima, this frontier of
points, is that they correspond to operating points where the
improvement of one objective function value comes only at the cost
of worsening some other objective function value---trading off one
for another.  Figure \ref{fig:paretofrontier} illustrates this set
of points in objective function space by the open {\sf O}'s where
each operational decision vector produces two objective function
values $f_1$ and $f_2$ that are minimized.  It is along this set
of points that operational decisions must be restricted if
operational efficiency is to be maintained. It is easy to see that
the Pareto frontier dominates (is superior to) the other points in
this 2-dimensional objective function space.

Effective methods for determining several Pareto optima can be
quite valuable for enhancing the survival value of a species (or
managing a complex system) because it enables {\em adaptive}
behavior. This allows the possibility of efficient operations when
there are changes in the relative utility of the several
objectives or when one objective must be restricted to a certain
range of values. Under such circumstances, the system can be moved
along the Pareto frontier from one Pareto optimum to another
thereby maintaining operational efficiency. Thus, if in an ant
colony a path to a food source becomes congested, then other
routes must be utilized. Although the distances to food sources
are generally minimized as is the level of congestion, these often
conflicting objectives can be efficiently traded off when the
shortest distance is sacrificed to lessen the level of congestion
\cite{SI-Bonabeau99}.

To summarize this first component of the foundational principles
of SI, the first principles and its implications suggest that both
subsystems and system should behave in a manner consistent with
Pareto optimality.  What remains is to characterize {\em Pareto
optimal behavior} in a context reflective of SI. It is noteworthy,
and somewhat surprising, that these ideas pertaining to Pareto
optimality and self-organization have apparently not been pursued
or even articulated by others in the context of SI research!  See
\eg the very complete books on SI such as \cite{SI-Bonabeau99}.
Assuming that natural selection pressures have imposed behaviors
leading to efficient, adaptable, \ie Pareto optimal behavior, the
question of how this is achieved in social insect societies must
be addressed. What are the mechanisms that social insects use to
solve what amount to MOPs with the limited forms of communication
they seem to have?

It may be that notions of scale are involved.  Indeed, the concept
of {\em stigmergy} suggests that what has been regarded as
collective intelligence or emergent behavior on a large scale is
effected by decisions on a smaller scale
\cite[p.14]{SI-Bonabeau99}. But how is such a relationship among
different scales possible when decisions affecting the local
environment are based on apparently {\em simple} information?  How
can system change be mathematically described in a way consistent
with movement along the Pareto optimal frontier?  More significant
however is the notion of {\em efficiently trading off the
efficiencies on one scale with efficiencies on another scale}.
Addressing these issues mathematically is the purpose of the
dynamical framework described below.

\subsection{The Dynamical Framework: Promoting
Adaptability}\label{sec:dynframe} The purpose of the dynamical
framework is to provide guidelines for how to mathematically model
the dynamics of the system in a way that is consistent with the
first principles and faithfully reflects the desired phenomenon.
Its articulation provides the necessary research and development
focus for describing system dynamics.  This allows {\em specific
questions to be asked\/} that lead to specific tests, mathematical
constraints, and the like.  In essence, it facilitates the {\em
engineering} of the dynamical rules that, in this case, lead to
Pareto optimal behavior. It is the essential feature that allows
this type of research to proceed.

For SI, the dynamical framework must be based on the two
intertwined principle features of SI---its emergent or
self-organized behavior, and the minimal information processing
and communications capabilities of the entities involved. These
two rather vague notions must somehow lead to more well-defined
properties and characterizations.  The preceding section suggests
that the emergent behavior can be characterized by Pareto optimal
operating points in some decision space.  The additional
requirement that such efficient behavior arises from limited
communication and information, in effect, imposes some scaling
aspects. These are addressed by the {\em Scale Invariant Pareto
Optimality} (SIPO) concept described in the next section.

\subsubsection{Scale Invariant Pareto Optimality}\label{sec:SIPO}

The SIPO concept stems from a belief that evolution has lead to
behaviors of social insects that promote the entire species.
Somehow over the eons of evolution, insects societies have
developed rules of conduct that are biologically driven and lead
to efficient management of their environment and resources
\cite[p.30]{SI-Bonabeau99}. Yet, given their apparently limited
capacity for communication and information processing, it seems
rather mysterious that in spite of these limitations, their
behaviors seem almost guided by an invisible hand\footnote{Their
economies like those of humans may indeed be guided by Adam's
Smith's {\em invisible hand}, \ie swarm intelligence!} that
somehow efficiently manages an entire colony's resources! How is
this possible?  As stated earlier, system movement along the
Pareto optimal frontier lends itself to adaptable and efficient
behavior thereby increasing the survival value of these social
insects. The heart of the matter in SI therefore becomes:
\begin{quote}
How can an entire system be moved along the Pareto optimal
frontier when the entities that comprise the system have only
limited information available to them on which to base their own
behaviors and where their behaviors have only limited effects on
their environment and other entities?
\end{quote}


The SIPO concept attempts to answer these types of questions.
Simply put, it is a way of characterizing Pareto optimal solutions
on many scales and thus reflects efficient and adaptable system
behaviors on many scales. It provides a basis for defining
properties of systems that are reasonable and have been observed
in social insect colonies.  Evolution can be seen as imposing not
just behaviors that enhance survival of individual entities, but
also of groups or societies of these entities
\cite[p.30-1]{SI-Bonabeau99}. Thus, while ants must manage their
own survival, they also derive distinct advantages from
participating in a colony. Indeed, scientists have observed
behaviors in ant colonies that in many respects is efficient on
several different scales involving sub-societies or sub-colonies
(see \cite[p. 209]{SI-Bonabeau99} for a discussion on nest
modularity).  There are divisions of labor, aspects of
specialization and such that contribute to the overall survival
value of an entire colony \cite[p.2]{SI-Bonabeau99}.  This
interdependency seems to promote their collective, and hence
individual, survival.

Since the SIPO concept revolves around the concept of Pareto
optima\footnote{It is worth noting that this description of SIPO
concepts seems, on the surface, to be similar to those articulated
by others. Menczer \cite{Menczer00} \eg describes an evolutionary
algorithm that seeks to determine Pareto optima and has some
scaling properties. But this article does not describe the
adaptable behavior of entities in terms of movement along a Pareto
frontier.  Coello \et \cite{Coello01} also provides an intriguing
title, but their paper also deals with evolutionary optimization
schemes for solving MOPs and does not address issues pertaining to
SI.} it is necessary to fully characterize its mathematical
attributes. So far, only its definition has been described. The
next few paragraphs describe another important aspect of Pareto
optima.

\paragraph{The Measure of Pareto Optima}\label{sec:MOPO}

A rather intuitive yet surprisingly little known aspect of Pareto
optima is its measure.  This measure is based on the size of the
set of points in objective function space that are dominated by
the Pareto optimal frontier---in essence a Lebesgue measure or
hypervolume. Zitzler \cite{Zitzler98} was apparently the first to
publish this idea in a brief three sentence
description.\footnote{This idea was independently formulated by
the author who later discovered that Zitzler had already described
it in a brief three sentence passage. The author was therefore
compelled to provide a formal proof and some algorithmic
extensions.} Fleischer \cite{Fleischer_EMO} extended this idea and
described a set function that maps a set of points in objective
function space to a scalar value, the Lebesgue measure, and
formally proved that this scalar achieves its maximum value if and
only if its arguments are Pareto optima. Fleischer also defined an
efficient (polynomial) algorithm for calculating this scalar for
an arbitrary number of dimensions (objective functions).

\begin{figure}[htb]
\vspace{-.1in}
\centerline{\includegraphics[width=2.4in]{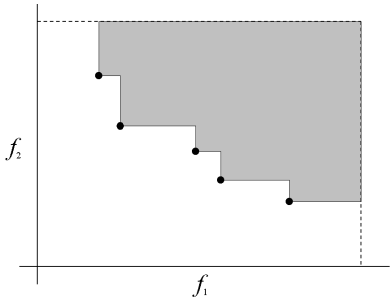}}
\vspace{-.1in} \caption{\small The hypervolume of Pareto optima.}
\label{fig:Lebesgue}
\end{figure}

This hypervolume is illustrated by the shaded region in
Figure~\ref{fig:Lebesgue} for two minimizing objective functions
$f_1$ and $f_2$ (see \cite{Fleischer_MOPO} for more complicated
illustrations in 3 dimensional space). The dotted lines are upper
bounds on $f_1$ and $f_2$ while the black dots correspond to
feasible points in the underlying decision space. Because this set
function produces a scalar, it can be used as an objective
function in an optimization algorithm such as SA. The proof that
if this hypervolume is maximized then its arguments are Pareto
optima\footnote{This direction of proof of the {\em if and only
if} implications is much more difficult, but is necessary in order
to justify its potential use as an objective function in a
metaheuristic such as SA where the mechanism of search is
crucially dependent on the objective function value, \eg the
Metropolis Acceptance Criterion the goal of which is to improve
the objective function value.} implies that its use in SA can
induce it to converge in probability to the Pareto
optima!\footnote{The author is not aware of any other
multi-objective optimization scheme that theoretically converges
in probability to Pareto optima.}

One important aspect of this measure is that it {\em quantifies
efficiency}, \ie changes in the overall efficiency can be
mathematically captured by a changes in the scalar hypervolume and
depicted by a receding or expanding Pareto optimal frontier (see
below for a further discussion on efficiency tradeoffs). This
measure and its potential roles in SI are described in the next
sections.

\paragraph{Subsystem Interactions and Constraints on
Objective Functions}\label{sec:interactions}

The SIPO property suggests that local efficiency may often be
sufficient for {\em system-wide} efficiency. Articulating and
applying this concept to systems of entities therefore imposes
constraints on how subsystems can interact, {\em using simple
forms of information}, so that a system, be it an ant colony, a
swarm of chemical sensors, or a network of computer systems,
operates on, or close to, the Pareto optimal frontier on both
local and global scales.  How this can be accomplished is
non-trivial because several identical subsystems may operate on
{\em different\/} Pareto optima in terms of local performance
metrics.

Consider the following simple example where two, identical
subsystems have operating points (vectors of parameters or
decision variables) on one of two local Pareto optima $\x_1$ and
$\x_2$. These two operating points lead to points in objective
function space $\p_1 = (f(\x_1),g(\x_1))\equiv (f_1,g_1)$ and

$\p_2 = (f(\x_2),g(\x_2))\equiv (f_2,g_2)$ for two local objective
functions $f$ and $g$, where, for purposes here, both are
minimized. Let functions $F$ and $G$ be global objective functions
each taking two arguments, one from each subsystem. Assuming the
order of arguments makes no difference in these functions, there
are three distinct ways in which these two subsystems can operate
on local Pareto optima: two ways where both subsystems operate on
the same point in local objective function space, \ie
$(F(\x_1,\x_1),G(\x_1,\x_1))$ and $(F(\x_2,\x_2),G(\x_2,\x_2))$
and one way in which the subsystems operate on different local
Pareto optima, \ie $(F(\x_1,\x_2),G(\x_1,\x_2))$. How should the
global objective functions $F$ and $G$ be defined in terms of
local objective functions so that all three combinations of local
Pareto optima map into the set of global Pareto optima?

More generally, under what circumstances do different {\em
combinations} of local Pareto optima associated with subsystems
yield a global Pareto optimum? Put another way, what is the form
of objective functions that allow mappings of Pareto optimal
points on the sub-system level, to Pareto optimal points on the
system-wide level?  This scaling from the sub-system level to a
system-wide level may therefore impose certain additivity
properties or exponential forms in the measures of performance and
presents separability issues.  For example, when must
$F(\x_1,\x_2) = f(\x_1)+f(\x_2)$? This is another example of how
the SIPO concept narrows down the playing field and sharpens the
questions that need to be asked and answered in any development
work involving SI concepts.

Although it is desired that subsystem states be maintained along a
local Pareto optimal frontier at the same time system-wide states
are maintained on a global Pareto optimal frontier, the
possibility exists that this may not always be possible depending
upon how these objective functions and their relationships are
defined. Afterall, subsystem states affect the system-wide
objective functions which are also affected by large numbers of
other subsystems. Moreover, subsystems do not behave
independently. The resulting complexity is similar in many ways to
the complexity of cellular automata (see below). The following
section describes how can these constraints can be captured in a
mathematically useful way that still reflects efficiency.

\paragraph{Efficient Tradeoffs of Efficiency Between
Scales}\label{sec:efficientTradeoff}

Sometimes the benefits to the individual entity must be traded off
for the good of the entire group---\eg soldiers must often make
the ultimate sacrifice to ensure the success of their unit (or
nation).\footnote{This type of behavior has also been observed in
insect colonies where soldier ants and worker bees sometimes
sacrifice their lives for the good of the entire colony. See
\cite[p.36-9]{SI-Bonabeau99}.}  This suggests the possibility that
certain states of neighboring subsystems maybe `incompatible'.
That is, a certain Pareto optimum in one subsystem may preclude a
particular Pareto optimum in another (similar to Ising spin glass
models---see the section below on Markov Random Fields). Thus, it
is conceivable that only certain combinations of Pareto optima
among neighboring subsystems may coexist at any given time. If
this is true, it may explain the notion of {\em fluctuations} as
Prigogine calls them \cite{Prigogine84} and the formation of
structures as an attribute of self-organization.

\begin{figure}[htb]
\vspace{-.1in}
\includegraphics[width=3in]{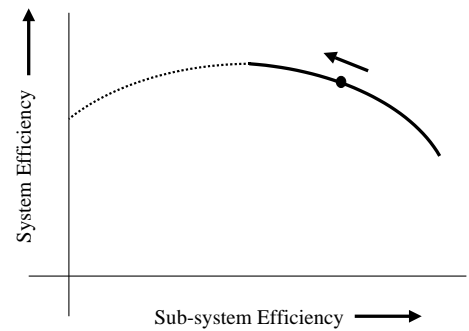}
\vspace{-0.3in}\caption{\small A tradeoff curve of different
measures of efficiency.}\label{fig:ParetoTradeoff}
\end{figure}

If such incompatibilities exists between and among neighboring
subsystems or between subsystems and system-wide Pareto optima,
then it must be the case that there is some tradeoff between the
efficiency measures of individual subsystems and/or the efficiency
measure of the entire conglomeration of entities. Some subsystems
may have to sacrifice efficiency in order for other subsystems to
operate Pareto optimally.  Recall from the earlier discussion that
the measure of Pareto optima is a quantification of the efficiency
of a system. Thus, a mathematical way of characterizing a tradeoff
of efficiency between two subsystems or between two scales of
systems is to use their respective measures of efficiency to
define a Pareto optimal frontier based on them. In effect, {\bf
the scalar hypervolumes associated with Pareto optima on different
scales themselves become objective functions for a Pareto optimal
frontier!\/} To the author's knowledge, this type of tradeoff has
never been characterized or described in this fashion, hence
provides a novel way of characterizing subsystem to system
interactions. Figure~\ref{fig:ParetoTradeoff} depicts such a
tradeoff curve where the Pareto optimal frontier is the bolded
part of the curve. Note that improvement on one scale comes at the
cost of worsening the hypervolume on the other scale. Note also
the concave form of the curve which is based on the ideas that
follow.\footnote{It should be borne in mind that the two axes
depicted, one for subsystems and the other for the entire system
are not independent, but this simple curve captures the essential
idea of the efficient tradeoffs of efficiency measures.}

\paragraph{Structures and Self-Organized Behavior}
The descriptions of self-organized behavior described earlier
often entail the notions of structures or patterns as the main
characterization of self-organization.  The foregoing discussion
on efficient tradeoffs of efficiency measures provides another way
of describing how such structures and their precipitating
fluctuations arise.

First, some random configuration among the subsystem or entities
occurs. Next, the subsystems all seek their local Pareto optimum.
Finally, in the case where some subsystems cannot attain a Pareto
optimum in terms of local objectives because of state
incompatibilities among neighboring subsystems, these subsystems
attempt an efficient tradeoff of efficiency measures as described
above.  These tradeoffs are between an estimate of the measure
associated with the global Pareto frontier and an estimate of the
measure associated with the local Pareto frontier.  Thus, as the
subsystems and system evolves, neighboring subsystems may evolve
in a way guided by these tradeoffs of efficiency measures. Thus,
the state incompatibilities ultimately give rise to structures as
subsystems attempt to operate on their local Pareto optimal
frontier. The patterns of states are therefore governed by their
initial conditions, the various state incompatibilities in terms
of Pareto optima, and the efficient tradeoffs of efficiency
measures in the midst of these subsystem state incompatibilities.
The patterns or structures observed in insect swarms can therefore
possibly be explained in these terms.

\paragraph{The `Graceful' Degradation of Performance}

The foregoing discussion also suggests a general and mathematical
way of describing the ``graceful degradation'' of system
performance. Such graceful degradation has been seen as an
important component in the management of large scale systems such
as the Internet \cite[Ch.5]{NRC02}.  Rather than suffering
catastrophic changes to a system, it may be possible for system
efficiency to be degraded in a more gradual manner. This is
because system efficiency can be measured using the Lebesgue
measure described earlier, hence the measures for different scales
can be traded off efficiently! This idea of efficient tradeoffs of
efficiency measures provides a {\em quantifiable guide} of how
best to achieve this sacrifice of efficiency on one scale for
improved efficiency on another scale. These types of issues can
arise in a variety of contexts.

For instance, during the operation of a complex system it may
happen that environmental changes, or changes in system state
results in a subsystem-system configuration depicted in
Figure~\ref{fig:ParetoTradeoff}. Here, the $x-$axis is an estimate
of a particular subsystem's efficiency measure and the $y-$axis is
an estimate of the global efficiency measure.  The black dot
corresponds to the subsystem/system efficiency point under current
environmental conditions.  These conditions suggest a tradeoff of
efficiency is possible where the system's efficiency can be
improved at a cost of decreasing the subsystem efficiency, \ie the
black dot is moved in the direction of the arrow.

Obviously, if the efficiency measures of enough subsystems are
degraded, it becomes increasingly difficult to compensate for
their degradation.  Eventually, the entire system's efficiency
measure will be reduced, but the actions of the subsystems in
establishing their movements along this Pareto optimal frontier of
these efficiency measures, provides for the most graceful
degradation of efficiency---because the degradation occurs
efficiently! How this occurs poses an interesting and, needless to
say, difficult problem owing to the many subsystem interactions
with the system, \ie the complexity problem and the fact that
these efficiency measures are not independent. Without be able to
quantify efficiency with this scalar hypervolume, however, it
would be impossible to address this problem of graceful
degradation in any mathematical way that provides some guidance
and direction for how to achieve this. This approach of
characterizing the subsystem to system efficiency tradeoff
therefore has many applications to both information networking,
understanding the economics and management of such systems, as
well as economics in general!

SIPO is a new way of articulating emergent behavior in the context
of systems operating under the imperatives of natural selection.
It provides a way to mathematically characterize properties of
systems on many scales and imposes constraints on how the
subsystems interact and functionally affect global measures of
performance. This makes implementation of efficient systems more
realistic, provides some underlying structure from which to
investigate mathematical forms for objective functions, has the
appealing feature of scale invariance and makes sense from an
evolutionary point of view. But can such properties arise in
systems with limited forms of communication and information
processing? The following section provides a clue, based on recent
research, that says it is indeed possible for Pareto optima to at
least be associated with limitations on the form of information.

\subsubsection{Limited Signalling Systems and Pareto Optimality}

The nexus of operational efficiency and SI seems to stem from how
the swarms of entities, \ie sub-systems, communicate and interact
with each other and their environment.  If we are to assume that
biological systems have evolved over time to operate efficiently,
then the signalling systems used by social insects must have some
mathematical connection to multi-objective optimization and the
concept of Pareto optimality.  Moreover, such a signalling system
must be simple in some sense. The SI paradigm as indicated in the
literature is based on the use of very simple (and apparently
low-cost) signalling methods. Study of ant colonies for example,
shows that entire social systems operate efficiently using simple
chemical signalling systems based on pheromones. These pheromones
constitute a conceptually simple signalling system and provide a
great deal of useful information, \eg temporal information since
the strength of the pheromone decays over time by the process of
evaporation (\cite[p.43]{SI-Bonabeau99}) in addition to memory of
where the ants have been. By interpreting these chemical signals
and responding to them with simple behaviors, ants perform complex
functions efficiently.

The idea of using a simple signal to convey information that leads
to efficient behavior is somewhat remarkable and the question
therefore arises as to whether a simple signal can even be
associated with Pareto optima.  The answer to this question is of
course that it is possible in the sense that {\em an entire set of
Pareto optima can be measured using a single scalar value}, the
hypervolume or Lebesgue measure as described earlier (in the
paragraph on the measure of Pareto optima).  Certainly, a single
scalar quantity qualifies as perhaps the simplest signal
\cite{Dorf01}. Thus, a strong connection between Pareto optima and
simple signals is possible. In effect, this scalar may constitute
a mathematical analogue of pheromone!
This scalar quantity not only provides the foundation for a whole
host of heuristic methods that can reveal Pareto optima (or near
Pareto optima, see
\cite{Fleischer_EMO,Fleischer_MOPO}\footnote{Note that these
results have also detailed {\em efficient algorithms} for
calculating the scalar associated with this set function for any
number of objective functions.}), but may also enable the
operational movement along a Pareto optimal frontier.



Pheromone may therefore be nature's way of mapping an entire set
of Pareto optimal points to a single scalar quantity and suggests
mathematical constraints on the form of functions that mimic
pheromone. Thus, just as maximizing entropy served as a guide in
developing Metropolis' equations of state \cite{Metropolis} that
ultimately lead to SA, maximizing this Lebesgue measure may serve
as a guide in developing appropriate signalling systems to mimic
pheromone {\em and how other related quantities should change over
time or in relation to other quantities}.  This is the heart and
purpose of the dynamical framework.

\paragraph{Swarm-based Solution Methods}

One important consideration in this discussion on Pareto optima,
efficiency and limited communications is how to tie it all
together.  How can one effectuate solution methods involving
simple signals that, in effect, solve MOPs?

One way alluded to earlier is to use the SA algorithm.  One of its
parallel forms, referred to as {\em cybernetic optimization by
simulated annealing} (COSA), uses scalars that are communicated
among processors to modulate the temperature control parameter
\cite{Fleischer_JOH,Fleischer_MIC97}.  These scalars provide a
feedback control system that improves SAs performance.  Like many
ants signalling each other with simple signals, parallel
processors running COSA communicate using simple signals in a way
that enhances their collective performance.

This form of parallel computation as many similarities to other
parallel forms such as Tabu Search and so-called ``random restart
local search'' methods \cite{Reeves93}.  These methods all share a
number of features in common with how swarms of insects solve
problems \eg by exploring many different paths simultaneously (see
\cite[p.55]{SI-Bonabeau99}). In particular, each processor in a
parallel system starts at some randomly selected point in decision
space and attempts to find the global optimum from that point.
With many such processors, the chances of finding the global
optimum in a reasonable amount of time increases.

\paragraph{Transformations and Movement Along the Pareto Optimal
Frontier}

An important aspect of multi-objective optimization problems is
that the vectors of decision variables that produce the Pareto
optima can be scattered throughout the decision space (the domain)
in an apparently random or haphazard manner
\cite{Fleischer_EMO,Fleischer_MOPO}.  This is due to the interplay
among the several objective functions---Pareto optima may lie on
the sides of hills rather than at their tops or bottoms (the local
minima or maxima). Thus, it is very difficult to determine where
these decision variable are in the decision space. Some
appropriate method to determine the Pareto optima and their
corresponding vectors of decision variables is necessary (see
\cite{Fleischer_EMO}).

It bears emphasis that two Pareto optima that are close to one
another in objective function space may have their underlying
vectors of decision variables quite widely separated and vice
versa.  Thus, assuming that a gradual change in objective function
space is desired, such as when there is a gradual change in the
relative utility of the different objective functions, some method
of moving from one vector of decision variables to another such
vector some distance away must be devised. In other words, once
the Pareto optima are identified, some method of transforming one
Pareto optimal decision vector to another is required.

One approach for performing such a transformation is through the
use of feedforward neural networks \cite[Ch.2]{Bar-Yam97}.  Such
networks can be trained so that a Pareto optimal decision vector
in the input produces the desired Pareto optimal decision vector
at the output.  Of course, a great deal of further research and
development in this context is necessary.

This dynamical framework based on the SIPO concept and its
relationships to emergent behavior, efficiency, and simple
signalling systems (as suggested by the Lebesgue scalar), provide
a framework that may enable novel approaches to SI research and
development in ways consistent with the first principles described
earlier. Together, the two main components of the SI paradigm,
self-organization as indicated by movement along a Pareto optimal
frontier and the minimal signalling systems as indicated by the
Lebesgue measure for Pareto optima, provide a mathematical
foundation from which fundamental theories and principles of SI
can be developed because they are more precisely defined and
characterized than the rather {\em ad hoc} approaches currently in
vogue that attempt to capture {\em some} aspect of SI.

To fully accomplish this potential however, requires some way of
applying these ideas to concrete and abstract problems, \ie test
bed problems that facilitate the testing, experimentation and
ultimately the application of these ideas. The problems used must
have some potential for exhibiting emergent phenomena and at the
same time allow for modelling the simple forms of interaction, the
stigmergy, that is the hallmark of SI. The {\em problem
framework}, the last of the three formalism components, provides
these important elements for SI research and development and is
described in the next section.

\subsection{The Problem Framework}\label{sec:problemframework}
An important aspect of the problem framework is that it allows
specific questions to be asked and problems explored. It must be
sufficiently general to not only capture the essential elements of
the phenomenon, but also offer some avenues on which to implement
the dynamical framework. It must also be specific enough to allow
useful and appropriate mathematical tools to be brought to bear.
The problem framework described here attempts to achieve this by
describing an abstract problem type based on what is termed here
as {\em swarming finite state machine} (SFSMs) models.

\subsubsection{Swarming Finite State Machine
Models}\label{sec:FSMmodels}

The idea behind SFSMs is that notwithstanding the limited
capabilities of swarms of entities, we should nonetheless provide
some way of flexibly describing their level of complexity.
Afterall, human societies present self-organized behavior and
certainly forms of `swarm intelligence'.  The allusion to Adam's
Smith's ``unseen hand'' (\cite{SmithAdam}) and Alvin Toffler's
social code based on society's propensity for concentration,
centralization, specialization, standardization, synchronization,
and maximization (see \cite{Toffler3}) suggest that the central
features of SI may be ubiquitous. Are these not attributes of very
complex forms of swarm intelligence? What is needed then is a way
to address a continuum of complexity and swarm intelligence.

SFSMs allow us to do this in that each finite-state machine (FSM)
has a finite set of inputs, outputs and internal states that
provide memory, the potential capacity for learning (where the
transitions to states are affected by learning methods--see
\cite{Sutton98}) and certainly the complexity that we hope to
manipulate and gain insight about.  Moreover, the level of
complexity can be changed by simply changing the numbers of states
that comprise the SFSMs.

SFSMs form a type of cellular automata that when coupled with
appropriate transformation functions and objective functions to
reflect operation along a Pareto optimal frontier will exhibit the
type of self-organized behavior defined in Section
\ref{sec:self-organization}. The states of each FSM will have some
functional dependence on neighboring FSMs and will be designed to
stabilize on efficient operating points. This may entail using
genetic algorithms (GAs) or involve neural networks to search the
space of transition functions or train on prescribed operating
points so as to induce the FSMs to operate on Pareto optimal
frontiers.

These types of simulations provide a real microcosm of the SI
world. It can effectively test whether natural selection pressures
are consonant with efficient operating points in the manner
described here. Fitness functions related to the efficiency
measures could be used to induce the SFSMs to achieve this. It
would indeed be interesting to see how these SFSMs can then be
designed (or evolved) to efficiently handle tradeoffs of
efficiency measures, a cornerstone of the SIPO concepts described
earlier.

The use of SFSMs allow for abstractions and, ultimately, practical
implementations.  The level of complexity can be controlled in the
sense that the size of the FSMs, its number of states, inputs and
outputs, can be decided upon {\em a priori}.  The state space of
the SFSMs, how its transformation functions are established and
how they interact of course require further research, but this
problem framework and the other two formalisms mentioned earlier
provide an important structure on which real swarm intelligence
can be studied and implemented for real-world problems.

It is worth pointing out, however, that much of what has been
described, the properties and characteristics and such, have
features in common with many other areas of inquiry in the
sciences and engineering.  The following areas describe some of
these features that offer some potentially useful tools and ideas
for developing SI systems.

\subsection{Other Scientific/Engineering Tools}

\subsubsection{Group Theory}\label{sec:grouptheory}
The structure of the emergent or self-organized behavior in SI
based on the meta-formalism, \ie movement along the Pareto optimal
frontier, provides a number of methodologic and analytical
advantages for exploring the fundamentals of SI. The most
significant aspect of this structure is that it may permit the use
of {\em group theory} in developing the fundamental theories of
SI. Group theory provides powerful mathematical tools as well as
the distinct possibility of developing novel insights.

This perspective requires a number of reasonable assumptions and
conceptions. A basic conception is that we can model each entity
$i$ in a swarm as a {\em subsystem} that exists in a number of
different {\em states} $\x_i = (x_{i1},\ldots, x_{in})$ defined by
$n$ parameters corresponding to values associated with the local
environment as well as the status of the agents themselves. The
social insects, the ants for example, can be viewed as the {\em
agents of change} in that they modify the states of these
subsystems. The states of each subsystem also affect the states
and agents of neighboring subsystems, hence is conceptually
similar to cellular automata described below.

The actions of these agents of change can be modelled as {\em
transformations} on the states of each subsystem.  Let $\x'$
denote the resulting state from the operation of a transformation
function $T_i$.  Now $T_i$ may, in effect, be a function of not
just the state of entity $i$, \ie $\x_i$, but also the states of
neighboring subsystems.  Thus, $T_i(\x_i) =
f(\ldots\x_{i-1},\x_i,\x_{i+1}\ldots)$ where the states of
neighboring subsystems are explicitly denoted and $f$ is a
vector-valued function. Subsequent states, those resulting from
the action of the agents of change can be related to the current
states by
\begin{displaymath}
\x_i' = f(\ldots\x_{i-1},\x_i,\x_{i+1}\ldots) = \hat{f}(\x_i)
\end{displaymath}
for some $\hat{f} \in \hat{F}_i$ that is functionally dependent on
the states of the neighbors of $i$.  Because of our first and
second formalisms, these transformations must restrict the output
states to the set of Pareto optima $P$ or near Pareto optima $P'$,
\ie $\x' \in S = P\cup P'$. In effect, the neighboring subsystems
select transformations from a family $\hat{F}_i$ of
transformations all of which map one point in $S$ onto another
point in $S$. Assuming the existence of identity and inverse
transformations for each such point, the set $\hat{F}_i$
constitutes a {\em symmetric group} (\cite{Rosen95}) over the
states $S$.

\subsubsection{Cellular Automata}\label{sec:cellularautomata}

Complexity theory, in particular theories regarding {\em cellular
automata} (CA) share a number of features with SI (see \cite[p.
245]{SI-Bonabeau99}). CAs are characterized by a set of {\em
cells} usually arranged in some geometric pattern (although this
is not technically required) and connected to other similar cells
by some neighborhood rule. Each cell has a finite number of states
which are determined by the states of its immediate neighbors.
State transitions are defined by simple rules, a central component
of cellular automata, that produce complex patterns often
described as self-organized \cite{Bar-Yam97,Chaudhuri97}. The
distinctions between SI and CA as proposed here involve the
connection between the emergent behavior and Pareto optimality.
CAs usually do not involve objective functions {\em per se}. It is
also worth noting that there is a deep connection between CA and
group theory \cite[Ch. 3: Group CA Characterization]{Chaudhuri97}.
Chaudhuri, \et \cite{Chaudhuri97} provide a number of theoretical
results on the connection between CA and cyclic groups and other
group properties.

It is also worth noting that the transformation functions that
determine how each cell evolves are usually very simple
functions---somewhat analogous to the limited forms of
communication associated with SI and stigmergy. But in SI, the
relationships among the entities is dynamic owing to the mobility
of the agents of change.  Entity relationships are not fixed as in
CAs. Notwithstanding this distinction, SFSMs described above can
model this mobility in a way consistent with a CA paradigm by
incorporating system states that model changing environmental
conditions.

\subsubsection{Multi-Objective Optimization}\label{sec:multiopt}
One important aspect in making a connection between
self-organization and Pareto optima is how these Pareto optima are
identified.  By creatively utilizing the Lebesgue measure, SA and
other heuristics can be engineered to search for these operating
points in a very direct way, by maximizing this Lebesgue measure.
This can be done using the COSA concept which happens to
incorporate simple signalling systems as well.  Many useful
analogies to SI thus become apparent.  Swarms of insects are
analogous to parallel processors, simple signals in SI are
analogous to the scalar Lebesgue measure or some estimate of it,
maximization of this Lebesgue measure is analogous to movement
along a Pareto optimal frontier, hence a type of self-organized
behavior. This also suggests an interesting, and perhaps profound,
unity of concepts between and among the various paradigms
described here.

\subsubsection{Game Theory}\label{sec:gametheory}
The dynamics of insect colonies involve competition and
cooperation among them and so unavoidably involve aspects of game
theory. The dynamics of competition and cooperation constitute the
major components of the more interesting aspects of game theory
\cite{Schelling60}. Game theoretic concepts such as the famous
{\em Nash Equilibrium} are really statements about Pareto
optimality and how interacting entities achieve the greatest
utility in a changing environment of competition and cooperation
\cite[p.193,339]{Luce85}.

\subsubsection{Markov Random Fields}\label{sec:MRFs}

The theory of {\em Markov Random Fields} (MRFs) shares a great
deal in common with that of Boltzmann Machines, SA and CA.  As
alluded to earlier in the paragraphs on {\em Efficient Tradeoffs
\ldots}), some subsystem states may not be allowed to coexist with
neighboring subsystem states. MRFs can model this type of behavior
and have the desirable Markov property in terms of spatial
attributes. This provides a mathematical structure to these
problems.  A large body of useful theoretical results on MRFs
exists (se \eg \cite{Bremaud99}) and may be quite useful in
addressing some of the issues in SI and how efficient tradeoffs of
efficiency measures may be analyzed.

\section{Conclusion}\label{sec:conclusion}
The growth of communications and networked systems imposes on a us
a need to explore new and imaginative ways to address the expected
problems in managing these systems.  Effectively utilizing the SI
paradigm requires some frameworks on which to first build the
relevant theory and then guide research toward practical
implementations. This article examined three aspects of a
meta-formalism that attempt to provide this necessary structure to
further research into SI and develop a more rigorous and
mathematically sound theory for its analysis.  The three elements
of the meta-formalism involve 1) a set of first principles based
on the laws of nature, 2) a dynamical framework, and 3) a problem
framework. The relevant laws of nature are those associated with
the theory of evolution and the effects of natural selection. The
theory of evolution provides the justification for describing
efficient behavior in social insect colonies.  This efficient
behavior can be mathematically characterized and associated with
Pareto optima. Thus, the self-organized behavior often associated
with SI can be given a more mathematical treatment when described
in terms of operating points that lie along a Pareto optimal
frontier. Operation along this frontier also provides a way of
mathematically describing adaptive behavior, one of the hallmarks
of SI.  These notions of efficiency and adaptability which are
implications of the first principles can be given effect as the
underlying concepts behind the second formalism component---the
dynamical framework.

The dynamical framework was described in terms of scale invariant
Pareto optimality (SIPO), a property that requires that the
behavior of entities within a swarm must be consistent with
efficient behavior and tradeoffs of efficiency on many scales.
SIPO thus provides a valuable set of conceptual constraints that
focusses the research efforts, provides ideas that can be
mathematically formalized and expressed, and features some of the
mathematical elegance of {\em self-similarity} so often found in
natural systems.

Finally, the problem framework provides a recipe for abstracting
the problem and, in effect, provides the modelling clay with which
to mold SI research so that it is consistent with the first two
formalism components. The problem framework employs the use of
swarming finite state machines models which provide a way to model
a continuum of complexity apparent in the SI exhibited by
different species.

In conclusion, it is the author's hope that this meta-formalism
and the ideas in this article will help facilitate research and
development using SI.  The resulting theories have potential for
application in many areas of inquiry involving complex systems.
The goal of making these increasingly complex systems more
efficient, scalable and autonomous will help to ensure their
effectiveness well into the future.

\section*{Acknowledgement}
The author was supported in part by the Center for Satellite and
Hybrid Communications Networks in the Institute for Systems
Research at the University of Maryland, College Park, and through
collaborative participation in the Collaborative Technology
Alliance for Communications \& Networks sponsored by the U.S. Army
Research Laboratory under Cooperative Agreement DAAD19-01-2-0011
and the National Aeronautics and Space Administration under award
No. NCC8-235.

\section*{Disclaimer}
The views and conclusions contained in this document are those of
the author and should not be interpreted as representing the
official policies, either expressed or implied, of the Army
Research Laboratory, the National Aeronautics and Space
Administration, or the U.S. Government.

\bibliography{FoundSwarm}\label{sec:references}

\begin{biography}[{\includegraphics[width=1.1in,height=1.6in,clip]{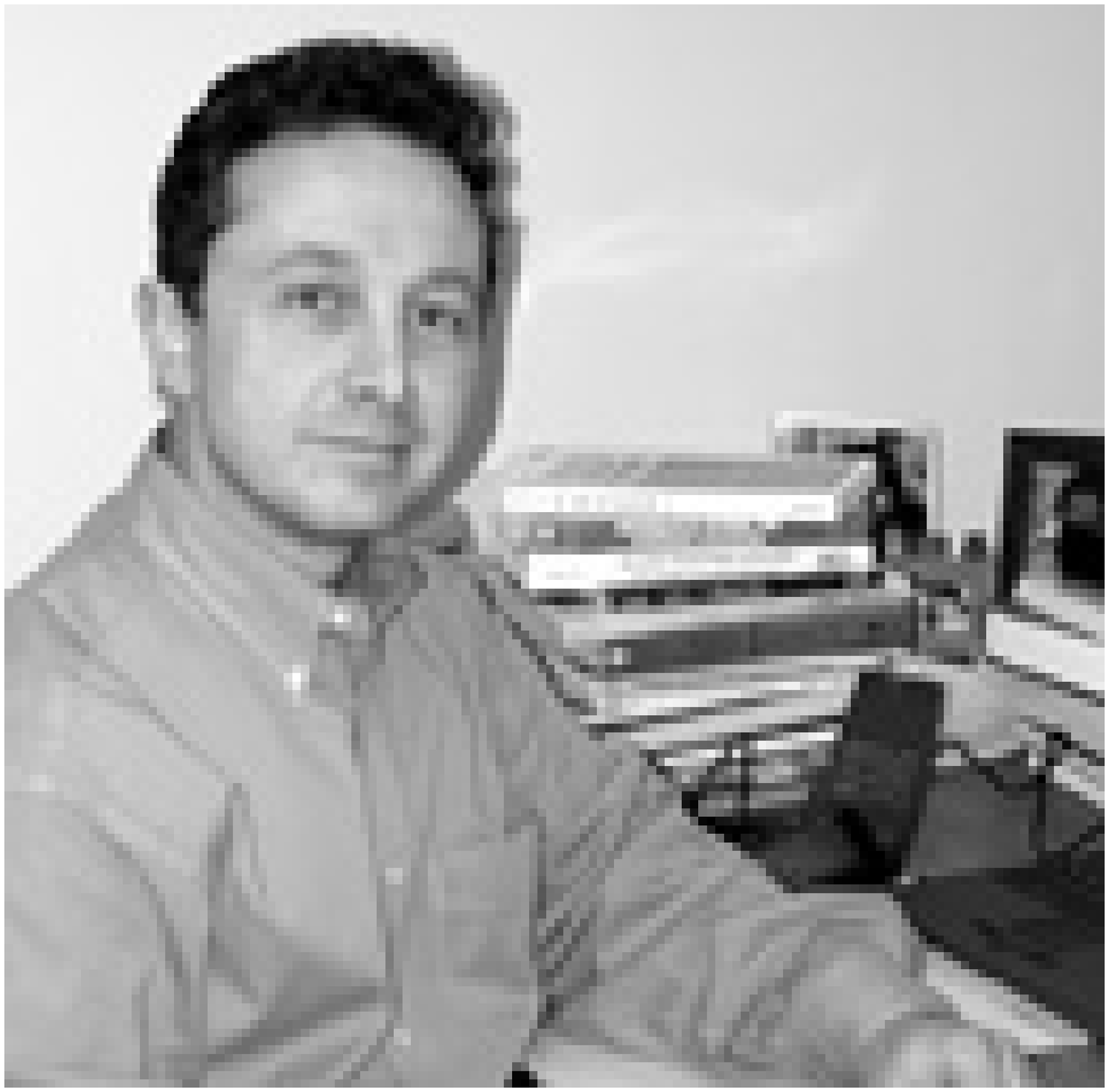}}]{Dr. Mark
Fleischer} is an Assistant Research Scientist in the Center for
Satellite and Hybrid Communications Networks at the Institute for
Systems Research at the University of Maryland, College Park. He
has a varied and multi-disciplinary background: a bachelor's
degree in political science from M.I.T., a law degree from
Cleveland State University, and a Ph.D. in operations research
from Case Western Reserve University. His interests span a number
of areas involving dynamical systems, stochastic processes,
applied probability, computer networking, and control theory. His
recent research involves optimization techniques in the domain of
multi-objective optimization where he developed results concerning
the measure of Pareto optima. His prior research involved
discerning the connections between the simulated annealing
optimization technique and information theory. He has also
developed new forms of parallelization techniques for simulated
annealing by using feedback control mechanisms to modulate the
algorithm's behavior. A recent paper of his describes his
discovery of scale invariant properties in the simulated annealing
algorithm and will soon appear in an upcoming journal article.
These optimization techniques have also been used in conjunction
with new and general resource allocation and scheduling models and
he continues to work on systems analysis, integration and decision
problems particularly in the area of swarm intelligence.
\end{biography}

\end{document}